# Strong coupling between a single quantum dot and an L4/3 photonic crystal nanocavity

Kazuhiro Kuruma[1,2]* †, Yasutomo Ota[3]*, Masahiro Kakuda[3], Satoshi Iwamoto[1,2,3] and Yasuhiko Arakawa[3]

[1]*Research Center for Advanced Science and Technology, The University of Tokyo, 4-6-1 Komaba, Meguro-ku, Tokyo 153-8505, Japan*

[2]*Institute of Industrial Science, The University of Tokyo, 4-6-1 Komaba, Meguro-ku, Tokyo 153-8505, Japan*

[3]*Institute of Nano Quantum Information Electronics, The University of Tokyo, 4-6-1 Komaba, Meguro-ku, Tokyo 153-8505, Japan*

* E-mail: kuruma@iis.u-tokyo.ac.jp, ota@iis.u-tokyo.ac.jp

† Present address: John A. Paulson School of Engineering and Applied Sciences, Harvard University, Cambridge, MA 02138, USA

**Abstract**

**We demonstrate strong coupling between a single quantum dot and a GaAs-based L4/3-type photonic crystal nanocavity. The L4/3 cavity supports a high theoretical $Q$ factor ($\sim 8 \times 10^6$), a small mode volume ($\sim 0.32\ (\lambda/n)^3$), and an electric field distribution with the maximum electric field lying within the host dielectric material, which facilitates strong coupling with a quantum dot. We fabricated L4/3 cavities and observed a high $Q$ factor over 80,000 using photoluminescence measurement. We confirmed strong coupling between a single quantum dot and an L4/3 cavity with a $Q$ factor of 33,000 by observing a clear anti-crossing in the spectra.**

Optical micro/nanocavities strongly coupled with semiconductor quantum dots (QDs) have been intensively investigated as a promising platform for the study of cavity quantum electrodynamics (CQED) in the solid state. Among the micro/nanocavities used for QD-based CQED studies, photonic crystal (PhC) nanocavities are one of the best resonators for exploring strong coupling regime because of their small mode volumes (*V*s) and high quality (*Q*) factors. Since the first observation of strong coupling between a QD and PhC nanocavity by measuring vacuum Rabi splitting (VRS)[1], strongly coupled QD-cavity systems have been utilized to observe various intriguing CQED phenomena[2–6]. The coherent light-matter interaction in the QD-cavity system is also essential for developing diverse photonic quantum information processing devices, such as nonclassical light sources[7,8] and quantum gates[9]. A figure of merit for strongly coupled QD-cavity systems is the ratio of the QD-cavity coupling constant ($\propto V^{-1/2}$), given by *g*, to the cavity decay rate ($\propto 1/Q$), *κ*. For pursuing better device performances and exploring novel physics, it is desirable to realize higher *g*/*κ*.

Considerable efforts have been devoted to improving *g*/*κ* in QD-based CQED systems, including the use of QDs with larger transition dipole moments[10,11], cavities with very small *V*s[2,12–14], and cavities with very high *Q* factors[15,16]. In our previous work, we experimentally demonstrated a very high *g*/*κ* of 6.4 with a large *g* of over 160 $\mu$eV using an H0-type PhC nanocavity with a high *Q* factor and a near-diffraction-limited *V*[17]. However, the electric field maxima of the fundamental resonance mode of the H0 nanocavity are located inside the airholes, hampering the maximization of *g*, hence limiting the achievable figure of merit. Recently, Minkov *et al.* theoretically proposed a novel cavity design[18], L4/3-type PhC nanocavity, formed by placing additional airholes in a PhC defect region, as an extension of the cavity design proposed by Alpeggiani *et al.*[19]. The fundamental mode of the L4/3 cavity supports a high design *Q* factor and a small *V*, both of which are comparable with those of the H0 nanocavity. Importantly, the field maximum of the L4/3 nanocavity mode is located at the cavity center filled with the host dielectric material. These properties are highly promising for achieving high values of *g*/*κ*. More recently, InP-based L4/3 nanocavities embedding InAs QDs have been experimentally demonstrated[20]. However, strong coupling between a single QD and an L4/3 nanocavity has not been reported.

In this study, we experimentally demonstrate strong coupling between a single QD and a GaAs-based L4/3 PhC nanocavity. The fundamental mode of the L4/3 nanocavity possesses a high design *Q* factor of ~8 million and a very small *V* of 0.32 ($\lambda$/*n*)$^3$. The L4/3 nanocavities were fabricated into a GaAs slab embedding InAs QDs and were investigated using micro-photoluminescence (*μ*PL) measurements. We observed a high experimental *Q* factor of over 80,000 for the fundamental cavity mode. With an L4/3 nanocavity exhibiting a moderate *Q* factor of 33,000, we observed strong coupling to a single QD with a clear VRS of 78 *μ*eV.

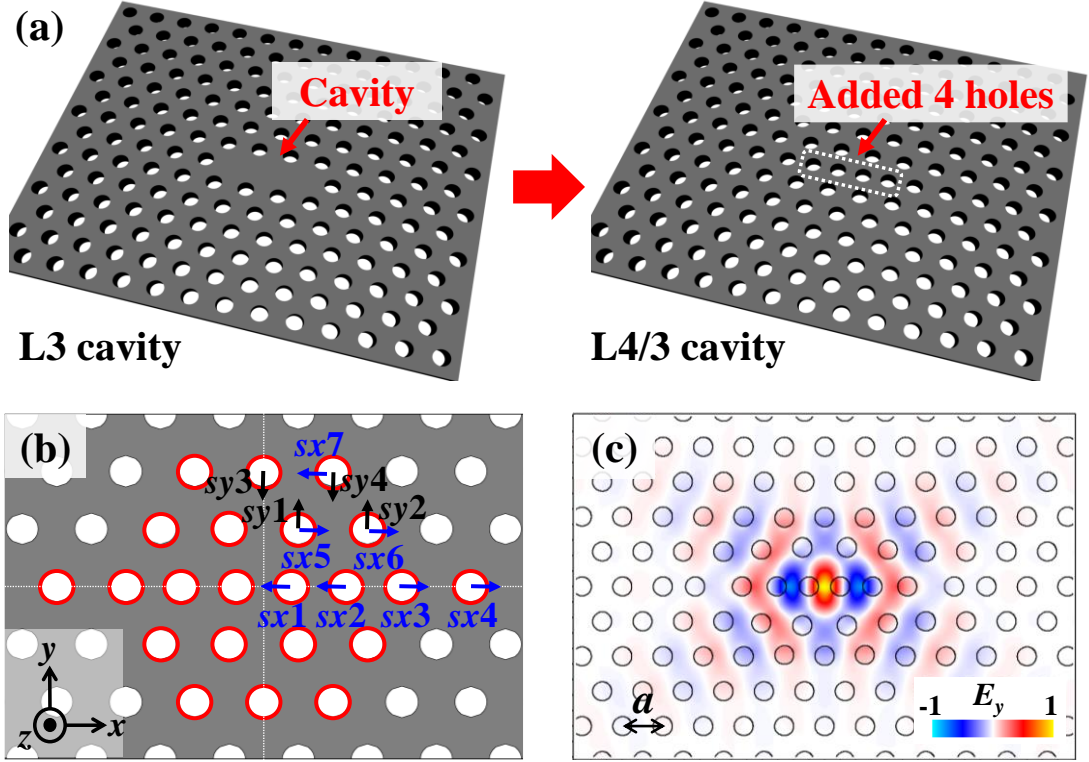

**Fig. 1.** (a) Schematic of an L4/3 (right) cavity formed from an L3 (left) cavity by placing additional four airholes in the cavity region (indicated as a white box). (b) Detailed description of the provided airhole shifts. The blue and black arrows indicate the shifts in the *x* and *y* directions, respectively. In the actual design, mirror-symmetric shifts were applied with respect to the plane across the cavity center (white lines). (c) Electric field distribution for the investigated fundamental cavity mode overlaid with the solid line indicating the airhole positions with the shifts defined in (b).

First, we numerically investigated an L4/3 PhC nanocavity with an air-suspended structure. The design started with an L3-type PhC nanocavity created by removing a row of three airholes in a hexagonal-lattice two-dimensional PhC. We added four airholes with equal intervals in the defect region to define an L4/3 cavity, as schematically shown in Fig. 1(a). The introduction of the additional airholes results in squeezing the resonance mode to the cavity center, leading to a larger reduction in $V$, compared to the original L3 design[18]. The PhC lattice constant ($a$), airhole radius ($r$), and slab thickness ($d$) were set to 260 nm, 61 nm, and 130 nm, respectively, to obtain the fundamental mode resonance at 970 nm (assuming the GaAs refractive index $n$ of 3.46). Then, for improving $Q$ factor, we introduced the shifts of the airhole positions in the $x$- ($sx_i$, $i = 1$–7) and $y$-directions ($sy_j$, $j = 1$–4), as illustrated in Fig. 1(b). By using the 11 shift parameters reported in [18], we obtained a very high calculated $Q$ factor of $\sim 8 \times 10^6$ while maintaining a very small mode volume $V$ of 0.32 $(\lambda/n)^3$. The latter value is comparable to the typical value of the H0-type nanocavity[17]. The reduction in the $Q$ factor compared to that in [18] ($Q \sim 20 \times 10^6$) seems to arise from the difference in the slab thickness. Figure 1(c) exhibits an electric field ($E_y$) distribution for the fundamental cavity mode calculated using the finite-difference time-domain algorithm. The electric field maximum is located at the cavity center within the dielectric material region, which enables maximum optical coupling to a single QD.

We compared the L4/3 nanocavity with other PhC nanocavity designs. Table 1 summarizes the key parameters of each cavity structure. The $Q$ values for all the cavities are over a million, which are high enough as designed $Q$ factors. This is because, in general, currently-achievable experimental $Q$ factors in active PhC nanocavities with QDs are much lower than their theoretical values due to fabrication imperfections, limiting the experimental $Q$ factors to $\sim 1 \times 10^5$[21] at best. Thus, we mainly discuss $V$s and resulting $g$s. $V$s of the L4/3 and H0 nanocavities are 3–4 times smaller than those of the others. In the Table 1, the impact of small $V$s is emphasized by including a column of the maximum-possible coupling constants ($g_{max}$) deduced using the simple relationship of $g \propto V^{-1/2}$. $g_{max}$s are normalized to that of the heterostructure cavity. $g_{max}$s for L4/3 and H0 designs are approximately two times higher than those for the L3 and heterostructure designs. For the H0 nanocavity, it could be difficult to locate a QD at which the electric field strength is more than 90% of the maximum due to the proximity of the etched sidewalls. In general, it should be avoided to locate QDs near the sidewalls, which could be a source of decoherence[22]. Therefore, among the two cavity designs realizing high $g$s, the L4/3 nanocavity is likely to be more advantageous for cavity QED studies. It should be noted that $V$ for the L4/3 cavity can be further reduced to $\sim 0.1$ $(\lambda/n)^3$ just by increasing the amount of shift $sx1$ (other airhole shifts need to be optimized for realizing a high $Q$ factor). However, this modification leads to a very narrow dielectric region near the cavity center, by which QDs are inevitably exposed to the influence of the side walls. In the current L4/3 cavity design, the minimum width of the dielectric region is ~80 nm, which is considered as sufficient to maintain a high coherence in the embedded QDs. We introduced double-periodic modulation[23] of airhole radii $\delta r$ by approximately ±1.0% around the cavity region to improve optical access to the actually fabricated nanocavities. This modulation reduced the theoretical $Q$ factor to ~6 × 10$^5$, while it barely changed the value of $V$.

**Table 1.** Comparison of the L4/3 nanocavity with other conventional PhC nanocavity structures. All values are theoretical. The second $g_{max}$ for the H0 cavity in parenthesis is a realistic value deduced by assuming that the QD is placed at the location where the cavity field intensity is 90% of its maximum, corresponding to a location ~30 nm from the airhole edge [17].

| Cavity type | $V\ (\lambda/n)^3$ | $Q\ (\times 10^6)$ | $g_{max}$ |
|---|---|---|---|
| L4/3 [18], [this work] | 0.32 | >8 | 2.2 |
| H0[24] | 0.25 | 1 | 2.4 (2.1) |
| L3[24] | 0.95 | 4.2 | 1.3 |
| Heterostructure[25] | 1.5 | 1580 | 1 |

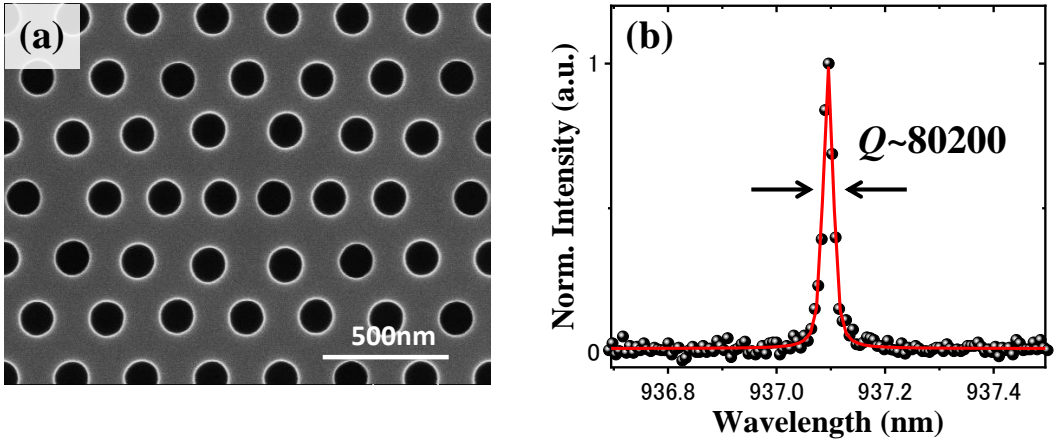

**Fig. 2.** (a) Top view of a scanning electron micrograph image of a fabricated L4/3 cavity with $\delta r$ of ±1.3%. (b) PL spectrum of the fundamental cavity mode. The solid red line is a fitting result using a Voigt peak function, whose Gaussian part was fixed to be our spectrometer resolution.

The designed cavity was fabricated into a 130-nm-thick GaAs slab grown on a 1-μm-thick $Al_{0.7}Ga_{0.3}As$ sacrificial layer, by a combination of electron beam lithography and reactive ion etching based on a chlorine-argon gas mixture and wet chemical etching using hydrofluoric acid. We also used a sulfur-based surface-passivation technique[21] to improve the $Q$ factors by reducing surface absorption. A single layer of InAs QDs, grown using molecular beam epitaxy with a low areal density of ~$10^8$ cm$^{-2}$, is contained in the middle of the slab. Emission peaks from individual QDs were observed around 930 nm. Figure 2(a) shows a scanning electron microscope (SEM) image of a fabricated L4/3 cavity taken after the wet etching process.

We performed μPL measurements at 6.2 K to optically characterize the fabricated samples, which were kept in a continuous flow liquid helium cryostat and pumped using an 808 nm continuous-wave laser diode. A 50× objective lens, with a numerical aperture of 0.65, was used to focus the laser light onto the sample. PL signals were collected using the same objective lens and analyzed by a spectrometer equipped with a Si CCD camera. The spectral resolution of the spectrometer was measured to be 21 $\mu$eV.

Figure 2(b) shows the PL spectrum of the fundamental cavity mode of an L4/3 cavity with $\delta r$ of ±0.9%. We used a low pump power of ~20 nW to avoid pump-induced degradation of $Q$ factors due to free carrier absorption[21,26]. By fitting the spectrum with a Voigt function, fixing its Gaussian part to the spectrometer resolution (the details of the fitting procedure can be found in Ref. [21]), we found a very high $Q$ factor of 80,200 (corresponding to a cavity decay rate $\kappa$ of 16 $\mu$eV). This value is smaller than the best value obtained using a heterostructure nanocavity with QDs[21]. The lower $Q$ factor may arise from the tighter light confinement in the L4/3 cavity, which makes the confined optical mode more sensitive to structural imperfections, as theoretically investigated in Ref. [18]. For further increasing $Q$ factors, we need to perform more careful optimizations of the fabrication process, especially around the cavity center region.

Next, we studied the optical coupling between a single QD and an L4/3 cavity. Figure 3(a) shows an emission spectrum of a sample taken under a far detuned condition. The peak at 936.73 nm is from the bare cavity emission, while the sharp peak at 937.36 nm is from a QD exciton transition. The $Q$ factor of the cavity mode was extracted to be 33,000 (corresponding to $\kappa$ of 40 $\mu$eV) from the fitted spectrum shown in the inset of Fig. 3(a). This $Q$ value is high enough to observe strong coupling with a single QD. The emission linewidth of the QD peak cannot be precisely evaluated as it is far below our spectrum resolution.

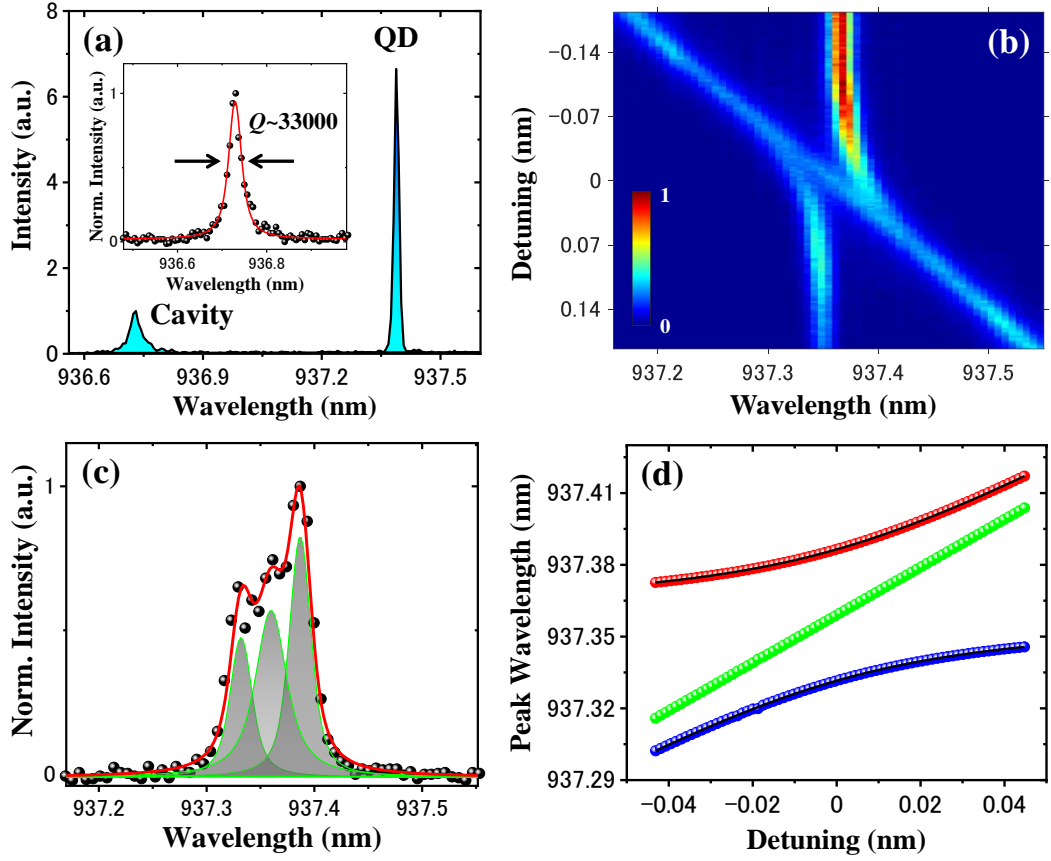

**Fig. 3.** (a) PL spectrum of a fabricated cavity with $\delta r$ of ±1.3%, coupled to a single QD. The inset shows an enlarged view of the cavity spectrum. The red line is the fitting curve of a Voigt function. (b) Color map of the PL spectra measured at 5 K under various detunings between the QD and cavity. (c) Vacuum Rabi spectrum taken at the QD-cavity resonance. The black dots indicate the experimental data. The solid red and light green lines indicate a fitting curve of multiple Voigt peak functions and individual peak components, respectively. (d) Peak positions of the upper (blue), lower (red) polaritons, and bare cavity (black) extracted by the fitting curve. Solid black lines are theoretically calculated polariton peak positions.

Then, we measured the detuning dependence of the emission spectra by shifting the cavity resonance frequency using a Xe gas condensation method[27]. Figure 3(b) shows a color map of the measured PL spectra when the cavity resonance was tuned across the QD emission line. We observed that the cavity and QD peaks do not cross each other around the QD-cavity exact resonance condition. This clear anti-crossing behavior suggests that the QD-cavity coupled system is in strong coupling regime. The antisymmetric PL intensities of the QD emissions between positive and negative detunings are perhaps induced by interactions between QD excitons and acoustic phonons[5]. Figure 3(c) shows a vacuum Rabi spectrum at zero detuning is plotted together with the fitting curves (light green lines). The two outmost peaks originate from the lower and upper cavity polariton states, while the center peak between them is from the bare cavity emission [2,13,28,29]. The bare cavity emission is often observed in QD-based cavity QED systems by pumping them non-resonantly [30]. Figure 3(d) shows a summary of the extracted peak positions of the polariton branches and the bare cavity mode by fitting with multiple Voigt peak functions. The plot again exhibits clearly the anti-crossing behavior of the system. From the separation between the two polariton peaks, the VRS was deduced to be 78 $\mu$eV. We also extracted a $g$ of 40 $\mu$eV from the splitting and $\kappa$ using the following equation: $g = \sqrt{(\text{VRS}/2)^2 + (\kappa/4)^2}$. Here, we neglected the small influence of the decoherence in the QD. The extracted peak positions in Fig. 3(d) also shown a good agreement with the theoretical calculations (solid black lines) using a standard CQED model [2]. Given the values of $g$ and $\kappa$, we obtained a ratio of $g/\kappa$ ~1, which safely fulfills the most primitive condition for entering strong coupling regime ($g > \kappa/4$). These results firmly demonstrate that the L4/3 cavity coupled to the QD was in strong coupling regime.

Finally, we discuss the possibility of realizing a higher $g$ using the L4/3 nanocavity. The obtained $g$ in this study is much smaller than that of ~165 $\mu$eV, previously reported using an H0 nanocavity with a similar $V$ of ~0.35 $(\lambda/n)^3$[17]. One reason for the reduced $g$ could be the large displacement of the QD position from its cavity field maximum. By considering our previously reported result of $g$ of 110 $\mu$eV, achieved with a position-resolved QD located near the field maximum (93%) of an L3 cavity with $V$ of ~0.75 $(\lambda/n)^3$[29], it is possible for the L4/3 cavity to obtain $g_{max}$ of ~180 $\mu$eV owing to the significant reduction of $V$. For positioning QDs to the field maximum, it would be necessary to use a precise QD position detection technique [2,29,31–33]. Another reason for the small $g$ could be the mismatch of the polarization between the QD dipole and the cavity local electric field. $g$ could be decreased by a factor of ~$1/\sqrt{2}$ because the cavity mode is linearly polarized, while the QD dipole is nearly circularly polarized in our QD, which confines a positively charged trion[34]. Hence, $g$ can be improved by using linearly polarized QD dipole emission from a neutral exciton and by aligning its dipole orientation with that of the local cavity field. Considering these possible improvements in the coupling strength, $g_{max}$ can reach to be ~260 $\mu$eV. We expect that a large figure of merit of $g/\kappa > 15$ is possible if using the L4/3 cavity with low $\kappa$ of 16 $\mu$eV, as we have already demonstrated in the device discussed in Fig. 2(b). The QD-cavity system with such high $g/\kappa$ will be important for the access to hitherto-unexplored QD-CQED physics, such as direct observation of the higher rungs of the Jaynes-Cummings ladder[35] in luminescence, as well as for the development of quantum information processing devices using photon blockade[7,8,36].

In summary, we have demonstrated a strongly coupled QD-nanocavity system using the L4/3 PhC nanocavity, which can possess a very high theoretical $Q$ factor of 8 million with a very small $V$ of ~0.32 $(\lambda/n)^3$. We performed $\mu$PL measurements to characterize the fabricated L4/3 cavities and observed a $Q$ factor exceeding 80,000. We also demonstrated strong coupling of a single QD to an L4/3 cavity with a high $Q$ factor of 33,000. Under the QD-cavity resonance condition, we observed a clear VRS of ~78 $\mu$eV, corresponding to an experimental $g$ of 40 $\mu$eV. Our results pave the way for realizing a QD-nanocavity system with a large $g$ and small $\kappa$, necessary for exploring previously-inaccessible CQED experiments and various QD-CQED devices that require highly coherent QD-CQED systems with large values of $g/\kappa$.

**Acknowledgments**

We would like to thank M. Nishioka and S. Ishida for their technical support. This work was supported by JSPS KAKENHI Grant-in-Aid for Specially Promoted Research (15H05700), KAKENHI (18J13565, 19K05300), JST PRESTO (JPMJPR1863), Iketani Science and Technology Foundation, Murata Science Foundation and a project of the New Energy and Industrial Technology Development Organization (NEDO).